\documentclass[aps]{revtex4}
\usepackage[T1]{fontenc}
\usepackage{graphicx}
\usepackage[latin9]{inputenc}
\makeatletter
\@ifundefined{textcolor}{}
{%
 \definecolor{BLACK}{gray}{0}
 \definecolor{WHITE}{gray}{1}
 \definecolor{RED}{rgb}{1,0,0}
 \definecolor{GREEN}{rgb}{0,1,0}
 \definecolor{BLUE}{rgb}{0,0,1}
 \definecolor{CYAN}{cmyk}{1,0,0,0}
 \definecolor{MAGENTA}{cmyk}{0,1,0,0}
 \definecolor{YELLOW}{cmyk}{0,0,1,0}
}

\makeatother

\begin{document}

\title{Low fertility rate reversal: a feature of interactions between  Biological and Economic systems}

\author{Jozef \v{C}ern\'{a}k}

\affiliation{Department of Nuclear and Sub-Nuclear Physics, Institute of Physics, Faculty of Science, Pavol Jozef Safarik University in Kosice, Ko\v{s}ice, Slovakia}

\begin{abstract}
An empirical relation indicates that an increase of living standard decreases the Total Fertility Rate (TFR), but this trend was broken in highly developed countries in 2005. The reversal of the TFR was associated with the continuous economic and social development expressed by the Human Development Index (HDI). We have investigated how universal and persistent  the TFR reversal is. The results show that in highly developed countries, 
$ \mathrm{HDI}>0.85 $, the TFR and the HDI are not correlated in 2010-2014. Detailed analyses of correlations and differences of the TFR and the HDI  indicate a decrease of the TFR if the HDI increases in this period. However, we found that a reversal of the TFR as a consequence of economic development  started at medium levels of the  HDI, i.e. $ 0.575<\mathrm{HDI}<0.85 $, in many countries.  Our results show a transient nature of the TFR reversal in highly developed countries in 2010-2014 and a relative stable trend of the TFR increase in medium developed countries in longer time periods. We believe that  knowledge of the fundamental nature of the TFR is very important for the survival of medium and highly developed societies.

\end{abstract}


\maketitle
\section{Introduction}
During the last decades, family life changed dramatically throughout the western world.  These changes are generally associated with the demographic transition \cite{Kirk}. Despite efforts to understand the transition, we  still lack clear comprehension \cite{Dribe_2014}. 

Population projections are notoriously imprecise \cite{Goodhart_1956, Lutz_2001, Lee_2011, Gerland, Abel_2016, Azose}. To understand the fundamental sources of this uncertainty it is useful to perceive the Total Fertility Rate (TFR) as a biological phenomenon \cite{Goodhart_1956}.  Goodhart \cite{Goodhart_1956} warned that ignoring the biological nature of fertility and population growth could cause serious issues. Diversity of the TFR decline in developing and developed countries \cite{Myrskyla} clearly shows that these warnings \cite{Goodhart_1956} are justifiable. This is one of the reasons why we consider natural means of population \cite{Goodhart_1956} to be the most important phenomenon to understand the TFR time series.  

In highly developed countries, the long-term declines of their TFRs have changed, and a new phenomenon, called TFR reversal, has been found and linked to continuous economic and social development \cite{Myrskyla}. The phenomenon gave rise to a certain optimism \cite{Tuljapurkar,Balbo, Burger2016} that has persisted until now. However, the development of the TFR after 2005, and specifically in the period of 2010-2014, shows no continuous  evidence of the TFR reversal in highly developed countries. 
 
In this study, the population is considered to be a living system \cite{Sneppen} that interacts with the environment \cite{Stulp}
and can be classified as a complex system \cite{Bar-Yam}. In such a system, interactions among its parts \cite{Cho2009} and environment could give rise to a broad spectrum of phenomena \cite{Cernak2016} for example, the demographic transition \cite{Kirk}, Lutz's assumption \cite{Balter1894} that countries with low fertility $\mathrm{TFR}<1.5$ may have crossed into permanent negative population growth, difficulties to project the TFR \cite{Lee_2011, Burger2016}, and etc.

Study of complex systems needs an interdisciplinary approach, where mathematical and physical methods are successfully used to understand phenomena like network formation, diffusion, collective behaviour, synchronization, phase transitions, chaotic behaviour, self-regulation, pattern formation, self-assembling, etc \cite{Bar-Yam, Cernak2016}. In the past, we invented and described fundamental properties of the digital generator of chaos \cite{Cernak_1996}, where an initial idea was to solve a nonlinear Feigenbaum equation on the lattice. Later, a similar approach was successfully used  to study the chaotic dynamics of experimental populations \cite{Henson_2001}. In Ref. \cite{Henson_2001} the authors  suggested  that such lattice effects could be an important component of natural population fluctuations because plants, animals and humans are counted in discrete units.  

A study of the relationship between the Human Development Index (HDI) and the TFR   provides  important information to propose a population model  that will take into account impacts of economic regulations and political changes on  individual woman's decision to plan a life path, as well as family size \cite{Cernak2016}.

\section{Data and methods}

The TFR \cite{WBank} is an average number of children that would be born to a woman over her lifetime. The HDI \cite{UNDP} is a summary measure of the average achievements in key dimensions of human development: a long and healthy life, being educated and having a decent standard of living. The HDI is a geometric mean of normalized indices for each of the three dimensions \cite{UNDP}.  

Time series of the TFR and HDI were downloaded from the portals of the World Bank \cite{WBank} and the United Nations Development Programme \cite{UNDP}. We have  analysed public data on HDI \cite{UNDP}, i.e no HDI coefficients have been re-calculated from the indices that HDI depends on. A disadvantage of this approach is that time series are short, however, it ensures a straightforward reproduction of the results. 

Spearman's rank correlation coefficients $\rho$ were calculated to test for correlations between the TFR and the HDI \cite{Myrskyla} in 1980, 1990, 1995, 2000, 2005, and 2010-2014. The  coefficients were determined in intervals of the HDI:   $0.0 \leq \mathrm{HDI} \leq 0.85$, and $0.85 \leq \mathrm{HDI} \leq 1.0$ to verify the correlations \cite{Myrskyla} previously reported. They were also computed  in additional intervals: $0.7 \leq \mathrm{HDI} \leq 0.85$, and $0.7 \leq \mathrm{HDI} \leq 1.0$. The rank correlation coefficients $\rho$ are shown in Table \ref{tab:spearman}. Open source software  and libraries (Python and Scipy) were used to compute the values of  $\rho$ and the statistical significance  $\mathrm{P}$ shown in Table \ref{tab:spearman}.

We have analysed short- ($\frac{dHDI}{dt}$, $\frac{dTFR}{dt}$, $\Delta\mathrm{HDI}_{i}$, $\Delta\mathrm{TFR}_{i}$) and long-term ($\Delta\mathrm{HDI}_{L}$ and $\Delta\mathrm{TFR}_{L}$) changes of the HDI and the TFR (note that the $t$ is time and $i$ is an index that will be defined in the next section).

\section{Results}\label{sec:reults}

\subsection{A transient behaviour of the TFR reversal in highly developed countries}

\subsubsection{Correlation of the HDI and the TFR}
\begin{figure}[h!]
\centering\includegraphics[width=8cm]{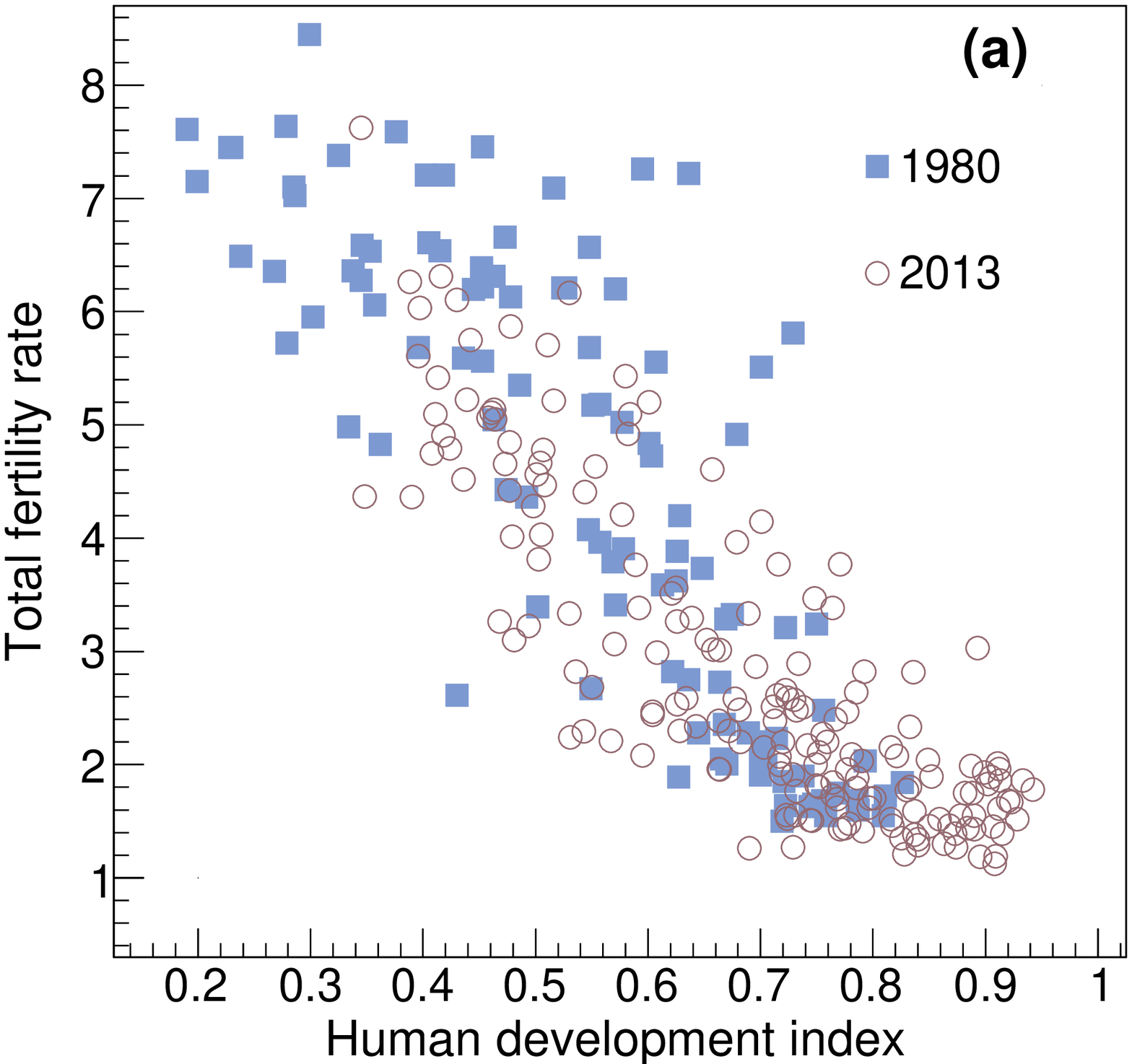}
\centering\includegraphics[width=8cm]{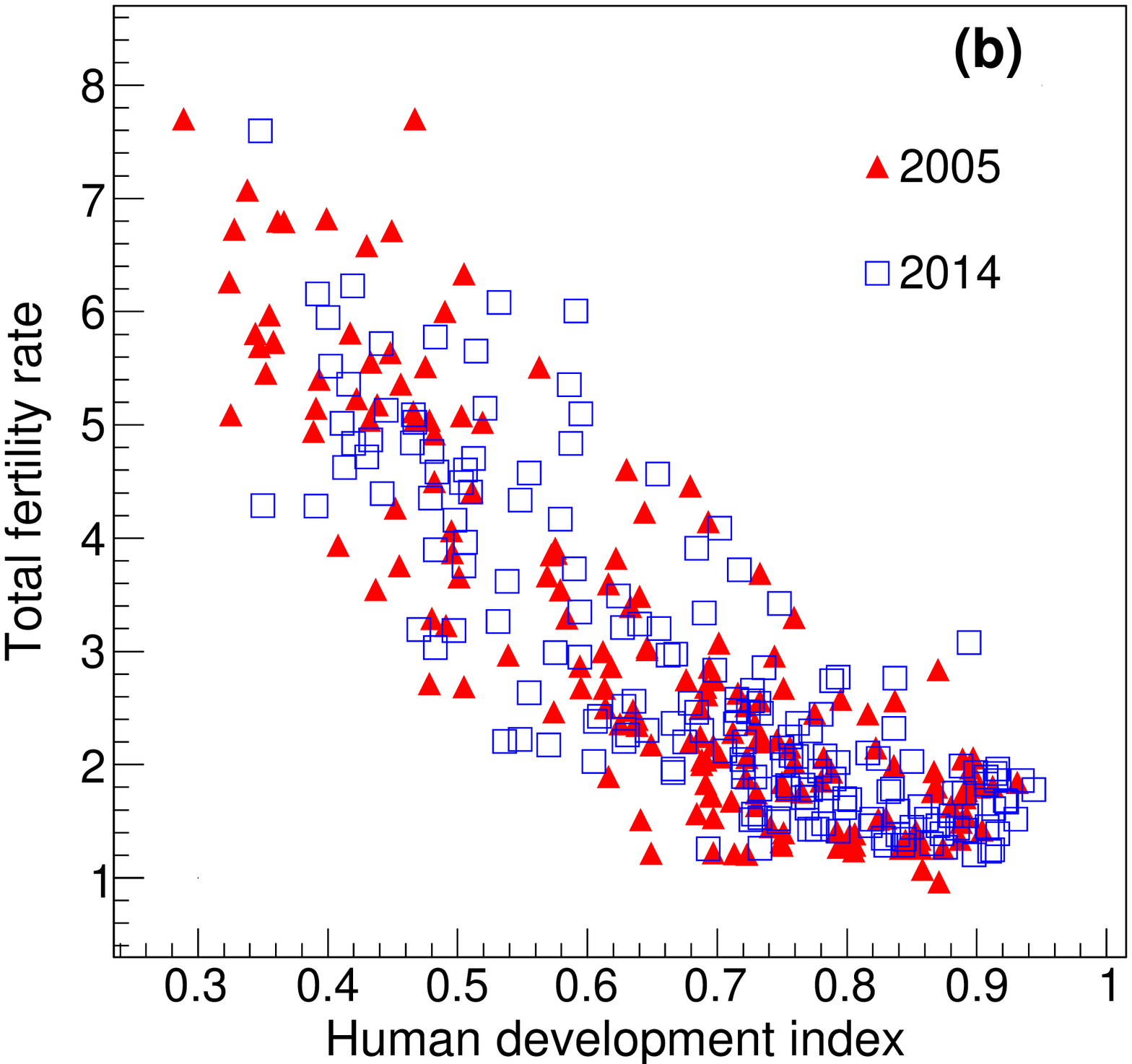}

\caption{The Total Fertility Rate (TFR) \cite{WBank} versus  the Human Development Index (HDI) \cite{UNDP} in (\textbf{{a}})  $1980$ and $2013$, then in  (\textbf{b})  $2005$ and $2014$.}
\label{fig:fig1}
\end{figure}

Graphs of the TFR  versus HDI per country for years 1980, 2005, 2014 and 2015 are shown  in  Figure \ref{fig:fig1}. These data were used to identify significant intervals of the HDI in which to test the correlation between the HDI and the TFR. Spearman's rank coefficients for these data  are shown in Table \ref{tab:spearman}.

In highly developed countries, $\mathrm{HDI}\geq 0.85$,  the TFR and the HDI were correlated in 2005 with the rank correlation coefficient  $\rho>0$ (Table \ref{tab:spearman}), which indicated the TFR reversal \cite{Myrskyla}. However, the coefficients $\rho$ in Table \ref{tab:spearman} show that before and after the year 2005, the TFR and the HDI were not correlated in these countries.

We found a new correlation interval, $0.7 \leq \mathrm{HDI}\leq 1.0$, with $\rho < 0$, that emerged in 2005 and has persisted until 2014. The TFR and the HDI are also correlated  in the interval $\mathrm{HDI} < 0.85$  with the coefficients $\rho <0$.

\begin{table}

\caption{\textbf{Spearman's rank coefficients $\rho$  with significance $\mathrm{P}$} are arranged in the format $ \rho (\mathrm{P})$. The values of $\rho$ are statistically significant  if $ \mathrm{P} <0.05 $ }

\begin{center}
\begin{tabular}{r @{\extracolsep{\fill}} rrrrr}
\hline 
$\mathrm{Year\  }$ & $\mathrm{HDI}<0.85$ & $\mathrm{HDI}\geq0.85$ & $0.7\leq\mathrm{HDI}\leq0.85$ & $0.7\leq\mathrm{HDI}\leq1.0$ \tabularnewline
\hline 
$1980\  $ & $-0.836\,(0.0)$ & - & - & - \tabularnewline
$1990\  $ & $-0.857\,(0.0)$ & - & - & - \tabularnewline
$1995\  $ & $-0.825\,(0.0)$ & $0.485\,(0.1850)$ & $-0.146(0.3260)$ & $-0.255\,(0.1170)$ \tabularnewline
$2000\  $ & $-0.830\,(0.0)$ & $-0.059\,(0.8200)$ & $-0.218\,(0.1600)$ & $-0.120\,(0.3630)$ \tabularnewline
$2005\  $ & $-0.845\,(0.0)$ & $0.500\,(0.0100)$ & $-0.308\,(0.0280)$ & $-0.288\,(0.0110)$ \tabularnewline
$2010\  $ & $-0.834\,(0.0)$ & $0.307\,(0.1100)$ & $-0.385\,(0.0010)$ & $-0.400\,(0.0001)$ \tabularnewline
$2011\  $ & $-0.836\,(0.0)$ & $0.308\,(0.0970)$ & $-0.374\,(0.0016)$ & $-0.364\,(0.0002)$ \tabularnewline
$2012\  $ & $-0.835\,(0.0)$ & $0.244\,(0.1780)$ & $-0.396\,(0.0008)$ & $-0.422\,(0.0001)$ \tabularnewline
$2013\  $ & $-0.815\,(0.0)$ & $0.217\,(0.2320)$ & $-0.412\,(0.0004)$ & $-0.457\,(0.0001)$ \tabularnewline
$2014\  $ & $-0.825\,(0.0)$ & $0.190\,(0.2800)$ & $-0.431\,(0.0002)$ & $-0.288\,(0.0002)$ \tabularnewline
\hline 
\end{tabular}
\end{center}

\label{tab:spearman}
\end{table}

In 2014, the TFR in highly developed countries (Table \ref{tab:TFRdecline}) is higher   than  the TFR in reference years (Table  S2 in Supplementary Material \cite{Myrskyla}), which is a signature of the TFR reversal in long time periods (see Section \ref{sec:londelta}).

Data of the HDI and the TFR in 2005 and 2014 in Table \ref{tab:TFRdecline} show  a transient decrease  of the TFR while the HDI increased in Norway, the Netherlands, the United States, Denmark, Spain, Luxembourg, Finland, Iceland and New Zealand. However, the TFR decrease differs from the predictions of Myrskyl\"{a} \textit{et al} \cite{Myrskyla}.

\begin{table}
\caption{\ The HDI and the TFR in 2005 and 2014. The list of highly developed countries is the same as in Table S2 of the Supplementary material of Myrskyl\"{a} \textit{at al} \cite{Myrskyla}.
}
\begin{center}
\begin{tabular}{c c c c c}
\hline 
 & \multicolumn{2}{c}{2005} & \multicolumn{2}{c}{2014}\tabularnewline
\hline 
Country & HDI & TFR & HDI & TFR\tabularnewline
\hline 
Norway & $0.931$ & $1.84$ & $0.944$ & $1.78$\tabularnewline
Netherlands & $0.891$ & $1.73$ & $0.922$ & $1.68$\tabularnewline
United States & $0.897$ & $2.05$ & $0.915$ & $1.86$\tabularnewline
Denmark & $0.902$ & $1.8$ & $0.923$ & $1.67$\tabularnewline
Germany & $0.887$ & $1.36$ & $0.916$ & $1.39$\tabularnewline
Spain & $0.845$ & $1.33$ & $0.876$ & $1.27$\tabularnewline
Belgium & $0.866$ & $1.72$ & $0.89$ & $1.75$\tabularnewline
Luxembourg & $0.88$ & $1.7$ & $0.892$ & $1.55$\tabularnewline
Finland & $0.869$ & $1.8$ & $0.883$ & $1.75$\tabularnewline
Israel & $0.87$ & $2.82$ & $0.894$ & $3.08$\tabularnewline
Italy & $0.856$ & $1.32$ & $0.873$ & $1.39$\tabularnewline
Sweden & $0.892$ & $1.77$ & $0.907$ & $1.89$\tabularnewline
France & $0.867$ & $1.92$ & $0.888$ & $1.99$\tabularnewline
Iceland & $0.889$ & $2.05$ & $0.899$ & $1.93$\tabularnewline
United Kingdom & $0.89$ & $1.8$ & $0.907$ & $1.83$\tabularnewline
New Zealand & $0.895$ & $2$ & $0.914$ & $1.92$\tabularnewline
Greece & $0.853$ & $1.28$ & $0.865$ & $1.3$\tabularnewline
Ireland & $0.895$ & $1.88$ & $0.916$ & $1.96$\tabularnewline
\hline 
\end{tabular}

\end{center}
\label{tab:TFRdecline}
\end{table}

\subsubsection{Differences of the HDI and TFR}

Using a longitudinal analysis as Myrskyl\"{a} \textit{et al} \cite{Myrskyla} did, we have found that $\beta^{pre}<0 $  and $\beta^{post}<0$,  where  $\beta^{pre}$and  $\beta^{post}$ are defined in \cite{Myrskyla}, i.e., no signature of the TFR reversal in the period of 2010-2014 is seen. The results are  not statistically significant and thus are not included in this report.     

Raw data \cite{WBank,UNDP} are arranged in time series of the $\mathrm{HDI}_{j,t}$ and the $\mathrm{TFR}_{j,t}$, where $j$ is a country index and $t$ is index of the year in 2010-2014.  We define the following sequences:
\begin{eqnarray}
\nonumber
\Delta\mathrm{TFR_{\mathit{i}}=TFR_{\mathit{j,t+1}}-TFR_{\mathit{j,t}}},\\ \nonumber
\Delta\mathrm{HDI_{\mathit{i}}=HDI_{\mathit{j,t+1}}-HDI_{\mathit{j,t}}},\\ 
\mathrm{HDI_{\mathit{i}}=
\left(HDI_{\mathit{j,t+1}}+HDI_{\mathit{j,t}}\right)/2},
\label{eq:hdi}
\end{eqnarray}
where $\Delta \mathrm{TFR}_{i}$ is a difference of the TFR,  $\Delta \mathrm{HDI}_{i}$ is a difference of the HDI, and $\mathrm{HDI}_{i}$ is an average value of the endpoints of the interval $\Delta \mathrm{HDI}_{i}$.

\begin{figure}
\centering
\includegraphics[width=11cm]{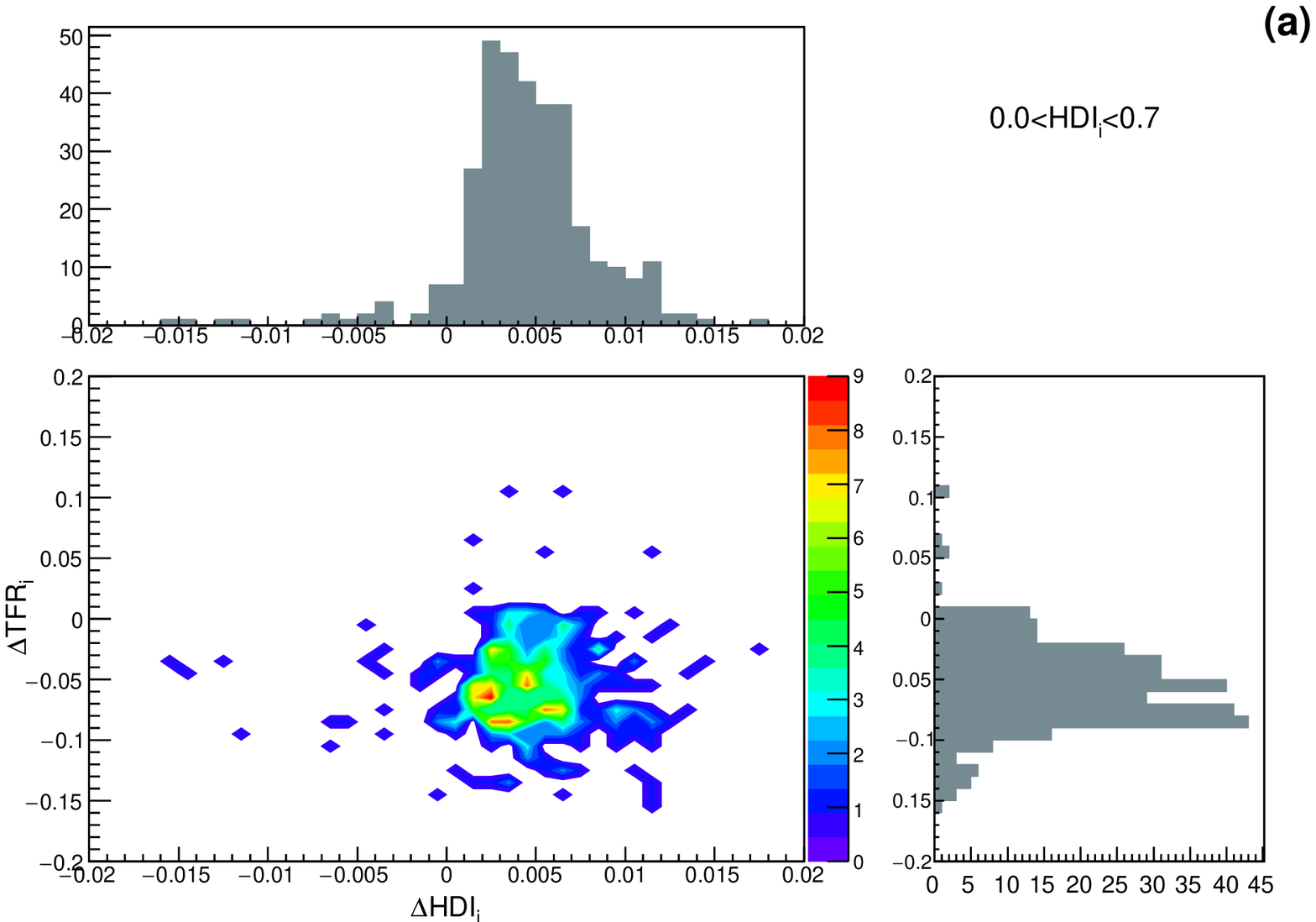}\\
\includegraphics[width=11cm]{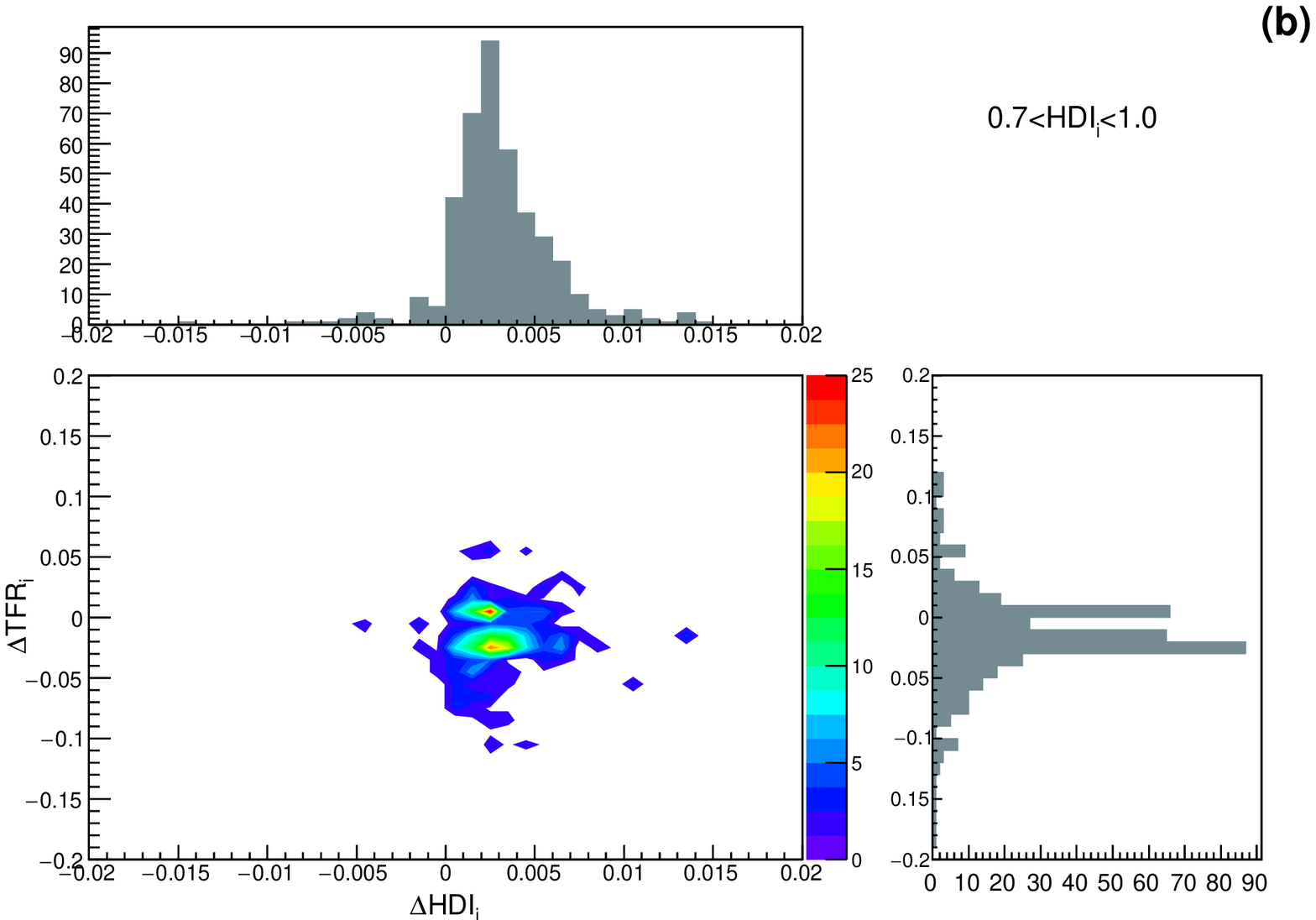}\\
\includegraphics[width=11cm]{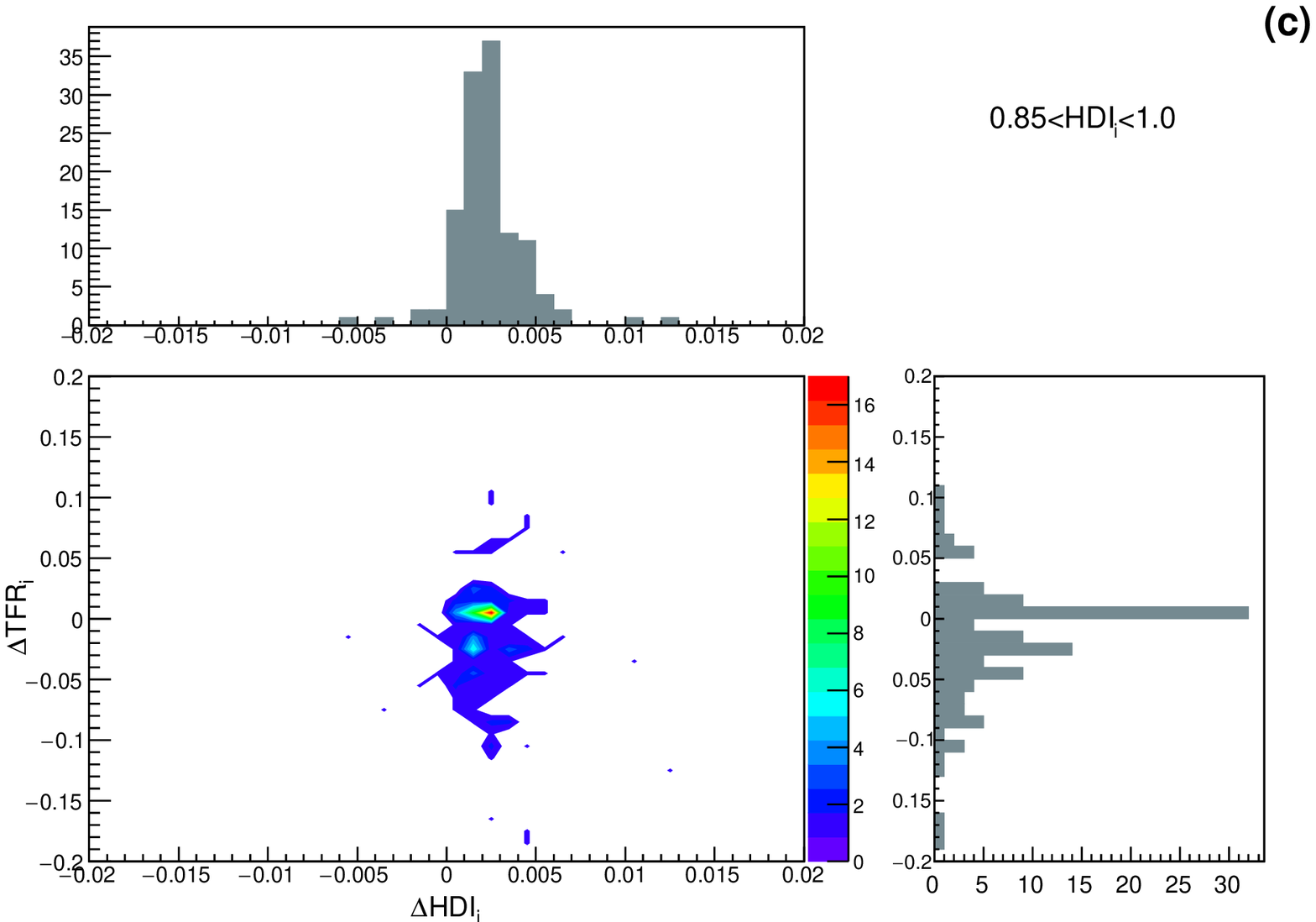}
\caption{2D histograms of differences of $\Delta \mathrm{HDI}_{i}$ and  $\Delta \mathrm{TFR}_{i}$ are shown as contour plots with the projections.  The differences of $\Delta \mathrm{HDI}_{i}$ and  $\Delta \mathrm{TFR}_{i}$  are analysed in  the specified intervals of $\mathrm{HDI}_{i}$: (a) $0.0<\mathrm{HDI}_{i}<0.7$, (b) $0.7<\mathrm{HDI}_{i}<1.0$, and (c) $0.85<\mathrm{HDI}_{i}<1.0$. The symbols $\mathrm{HDI}_{i}$,  $\Delta \mathrm{TFR}_{i}$, and  $\Delta \mathrm{HDI}_{i}$ are defined in Eq. \ref{eq:hdi}.}
\label{fig:fig2}
\end{figure}

Two-dimensional (2D) distributions of  $\Delta \mathrm{TFR}_{i}$ vs $\Delta \mathrm{HDI}_{i}$  are plotted as contour plots  and projections (Figure \ref{fig:fig2}). The TFR,  HDI and their differences $\Delta \mathrm{TFR}_{i}$ and $\Delta \mathrm{HDI}_{i}$ are not correlated, thus the plots of the differences provide initial information about system dynamics.  The projections of  $\Delta \mathrm{TFR}_{i}$ and $\Delta \mathrm{HDI}_{i}$ (see Figure \ref{fig:fig2}) are asymmetric distributions, and distributions of $\Delta \mathrm{HDI}_{i}$ show  long tails in one direction.  The average of $\Delta \mathrm{HDI}_{i}$ is positive  and the average of $\Delta \mathrm{TFR}_{i}$ is  negative  in all cases (Figure \ref{fig:fig2}).  The average of $\Delta \mathrm{TFR}_{i}$ (Figure \ref{fig:fig2}) shows a tendency of absolute value decrease if the $\mathrm{HDI}_{i}$ increases. For example, compare Figure \ref{fig:fig2}a and Figure \ref{fig:fig2}c, where $\Delta \mathrm{HDI}_{i}<0.7$ and $0.85<\Delta \mathrm{HDI}_{i}<1$. 

The differences $\Delta \mathrm{TFR}_{i}$ and $\Delta \mathrm{HDI}_{i}$ are not correlated (Figure \ref{fig:fig2}).  This is one of the reasons  to introduce a new sequence of differences, $\Delta \mathrm{TFR}_{k}=\Delta \mathrm{TFR}_{i}B$, where $B=1$  if $\mathrm{HDI}_{i} > \mathrm{HDI}_{C}$,  otherwise $B=0$. We have used differences $\Delta \mathrm{TFR}_{k}$ to compute the average:
\begin{equation}
\langle\Delta\mathrm{TFR}_{k}\rangle=
\left(1/l\right)\sum_{k=1}^{l}\Delta\mathrm{TFR_{\mathit{k}}}.
\label{eq:tfr}
\end{equation}

The average $\langle\Delta\mathrm{TFR}_{k}\rangle$ defined in Eq. \ref{eq:tfr} indicates an effect of small changes of the TFR, i.e $\Delta\mathrm{TFR_{\mathit{k}}}$, and the critical value of the HDI, i.e. $\mathrm{HDI}_{C}$, on the TFR.

\begin{figure}

\centering
\begin{center}
\includegraphics[width=8cm]{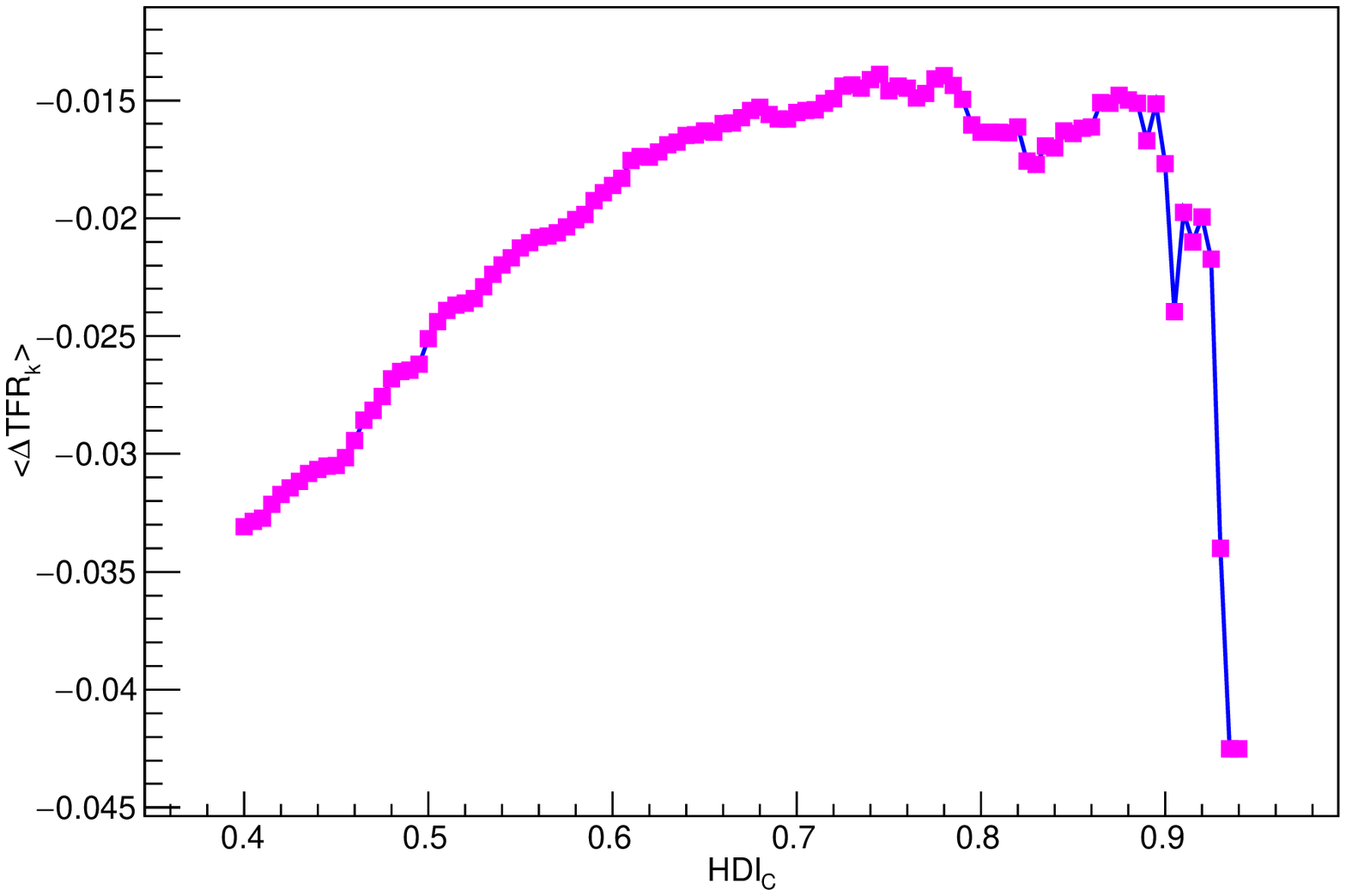}
\end{center}
\caption{The average changes in Total Fertility Rate  $\langle\Delta \mathrm{TFR}_{k}\rangle$  (see Eq. \ref{eq:tfr}) versus  the critical value of the Human Development Index $\mathrm{HDI}_{C}$. For $\mathrm{HDI}_{C}>0.9$ the average value is computed from less than $60$ entries of the $\Delta \mathrm{TFR}_{k}$. }
\label{fig:meas}
\end{figure}

We have analysed the time series of the TFR  in 2010-2014 using the differences $\Delta \mathrm{TFR}_{k}$ and the measure $\langle\Delta \mathrm{TFR}_{k}\rangle$  (Eq. \ref{eq:tfr}). The results are shown in Figure \ref{fig:meas}, where the absolute value of the average $|\langle\Delta \mathrm{TFR}_{k}\rangle|$ decreases if $\mathrm{HDI}_{C}$ increases up to the $\mathrm{HDI}_{C}\approx0.74$, then the graph shows a complex dependence of  $\langle\Delta \mathrm{TFR}_{k}\rangle$   on the critical $\mathrm{HDI}_{C}$, i.e., positive and negative effects of differences $\Delta \mathrm{TFR}_{k}$ on the TFR.  In the case of the TFR reversal which was found by Myrskyl\"{a} \textit{et al} \cite{Myrskyla},  the function  $\langle\Delta \mathrm{TFR}_{k}\rangle$ is an increasing function of the  $\mathrm{HDI}_{C}$ for  $\mathrm{HDI}_{C}>0.85$ and $\langle\Delta \mathrm{TFR}_{k}\rangle $ is greater than null for  $\mathrm{HDI}_{C}>0.85$, i.e., $\langle\Delta \mathrm{TFR}_{k}\rangle >0 $. 

\subsection{A reversal of the TFR in medium developed countries with $0.58<\mathrm{HDI}<0.85$ }

\subsubsection{Dynamics of the TFR changes}
\begin{figure}[h!]

\centering\includegraphics[width=8cm]{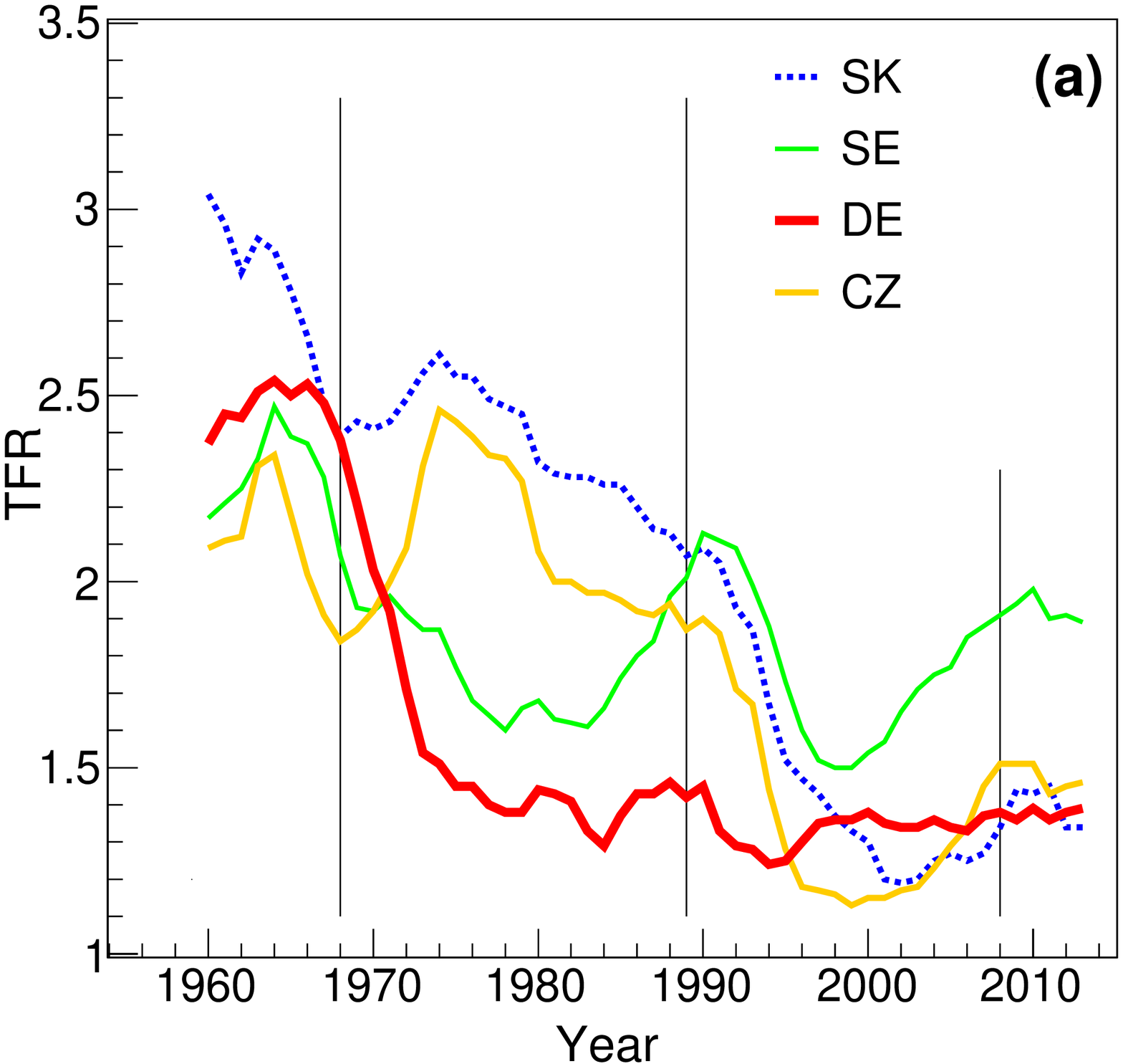}
\centering\includegraphics[width=8cm]{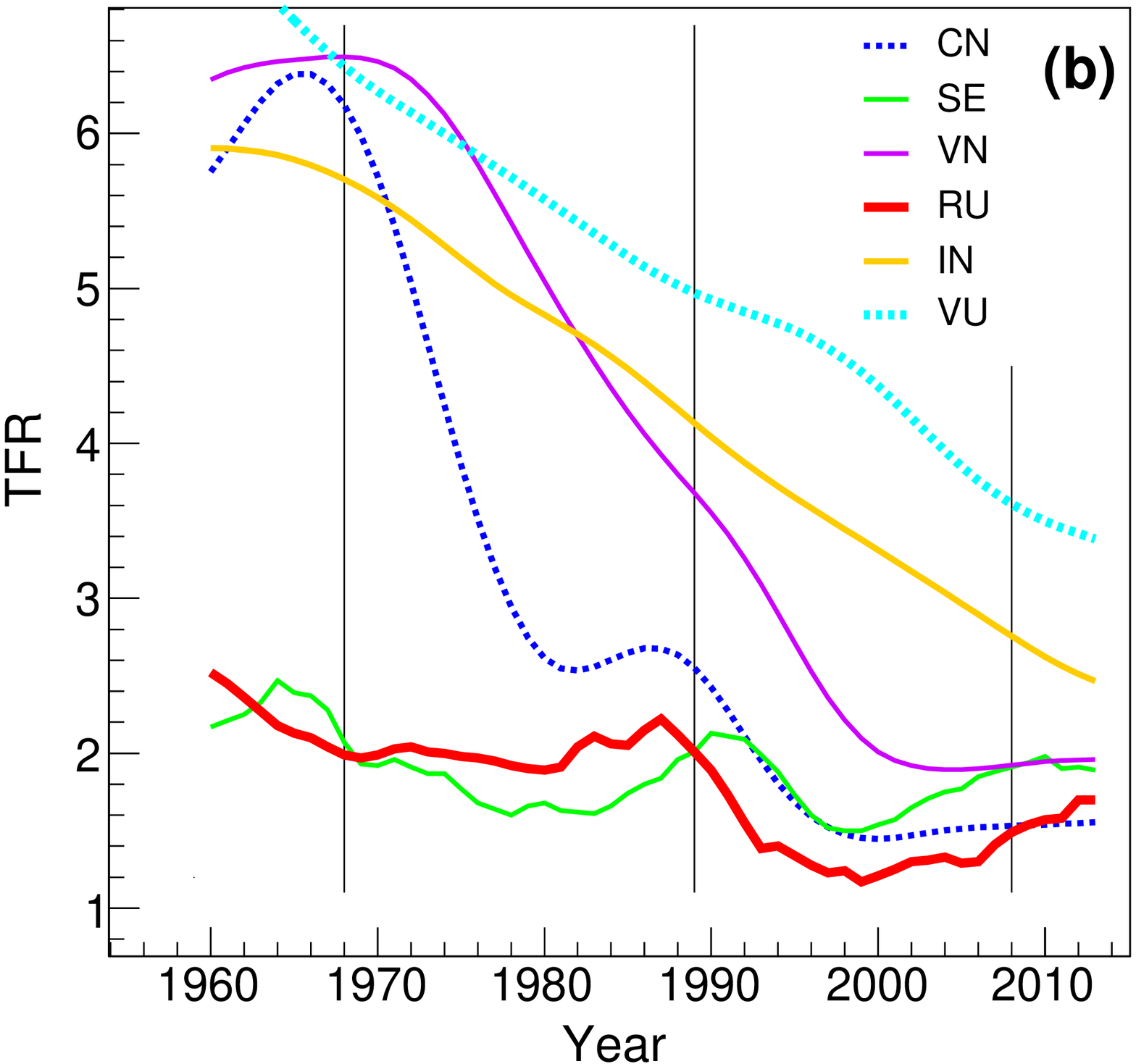}
\caption{The time series of the TFR in (a) Slovakia (SK), Sweden (SE), Germany (DE), Czech Republic (CZ), and (b) China (CN), Sweden (SE), Vietnam (VN), Russian Federation (RU), India (IN) and Vanuatu (VU). Vertical lines show important historical events: the energy crisis in 1970, the early recession in 1990, and the financial crisis of 2008.}
\label{fig:TFrtimeser}
\end{figure}

The time series in Figure \ref{fig:TFrtimeser} show that in certain time periods derivatives of TFR with respect to the time $t$, $\frac{dTFR}{dt}$,  are approximately equal (note that derivatives are not plotted in  Figure \ref{fig:TFrtimeser}). For example, in Figure \ref{fig:TFrtimeser}a, if TFR decreases, i.e., $\frac{dTFR}{dt} < 0$, the derivatives are approximately equal in Slovakia (SK) and Czech Republic (CZ) in 1965-1968, in Slovakia (SK) in 1965-1970 and in Germany (DE) in 1968-1973, and in Slovakia (SK), Czech Republic (CZ), and Sweden (SE) in 1992-1994, where the corresponding differences of the HDI in 1992-1994 were: $\Delta\mathrm{HDI}_{SK}=0.01$, $\Delta\mathrm{HDI}_{CZ}=0.015$, and $\Delta\mathrm{HDI}_{SE}=0.028$. In the case of  $\frac{dTFR}{dt} > 0$ the derivatives in Sweden (SE) and Slovakia (SK) are approximately equal in the period of 2002-2008, where $\Delta\mathrm{HDI}_{SK}=0.016$ and $\Delta\mathrm{HDI}_{SE}=0.049$. In Figure \ref{fig:TFrtimeser}b a decline of the TFR  in China (CN), Sweden (SE), Vietnam (VN) and Russian Federation (RU) show approximately the same derivatives $\frac{dTFR}{dt}<0$ in the period of 1990-1998, where $\Delta\mathrm{HDI}_{CN}=0.075$, $\Delta\mathrm{HDI}_{SE}=0.052$, $\Delta\mathrm{HDI}_{VN}=0.085$, and $\Delta\mathrm{HDI}_{RU}=-0.03$. In India (IN) and Vanuatu (VU) derivatives $\frac{dTFR}{dt}<0$ are approximately equal in the long period of 1970-2014, and for available data of the HDI in 2005-2015, $\Delta\mathrm{HDI}_{IN}=0.079$ and $\Delta\mathrm{HDI}_{VU}=0.026$, i.e. positive change of the HDI. Increases of the TFR in Sweden (SE) and Russian Federation (RU) show similar derivatives $\frac{dTFR}{dt}>0$  in the period of 1998-2010, where $\Delta\mathrm{HDI}_{RU}=0.082$ and $\Delta\mathrm{HDI}_{SE}=0.034$.

\subsubsection{Correlation of differences $\Delta\mathrm{HDI}_{L}$ and $\Delta\mathrm{TFR}_{L}$ \label{sec:londelta}}

\begin{figure}[h!]

\centering\includegraphics[width=8cm]{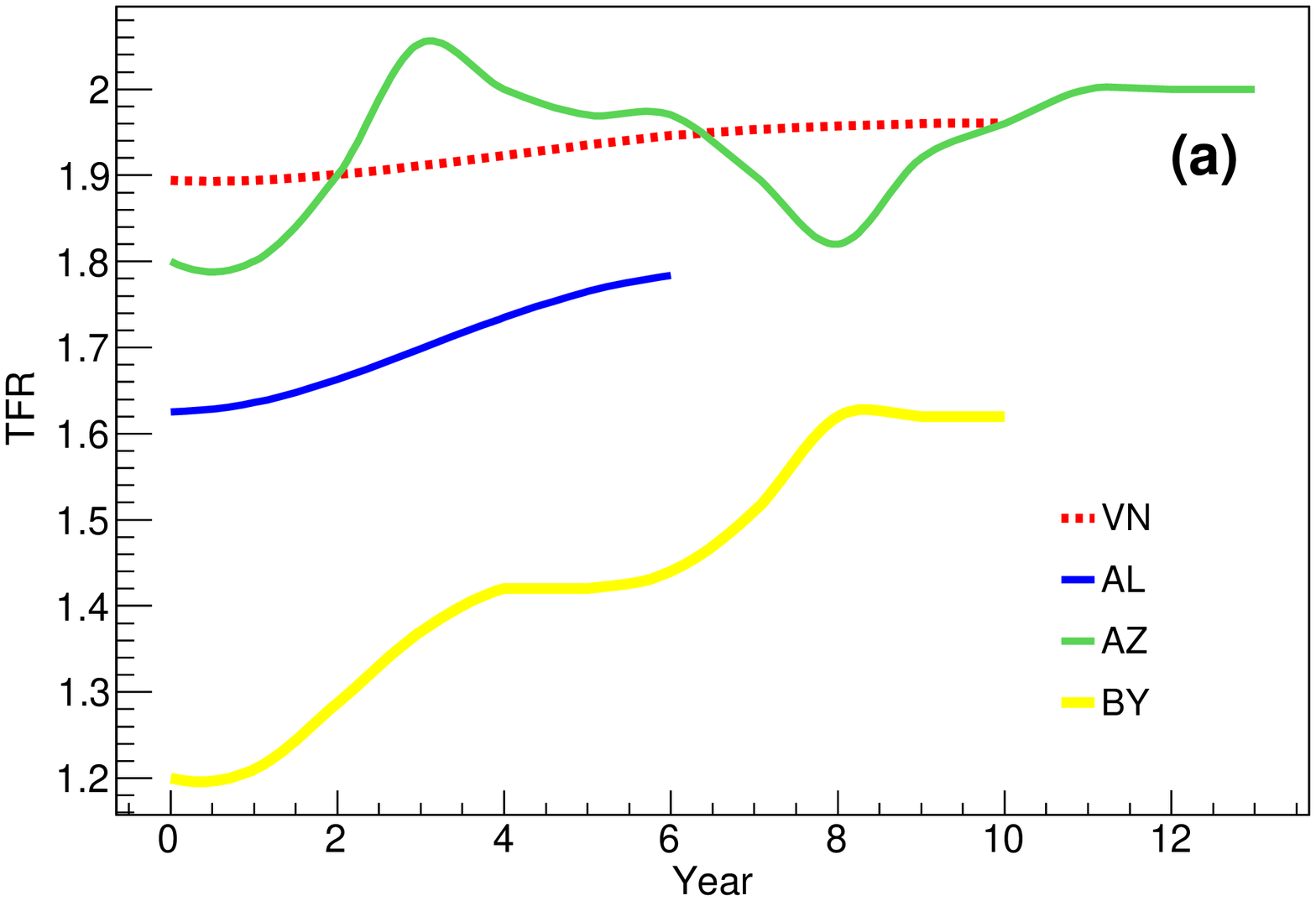}
\centering\includegraphics[width=8cm]{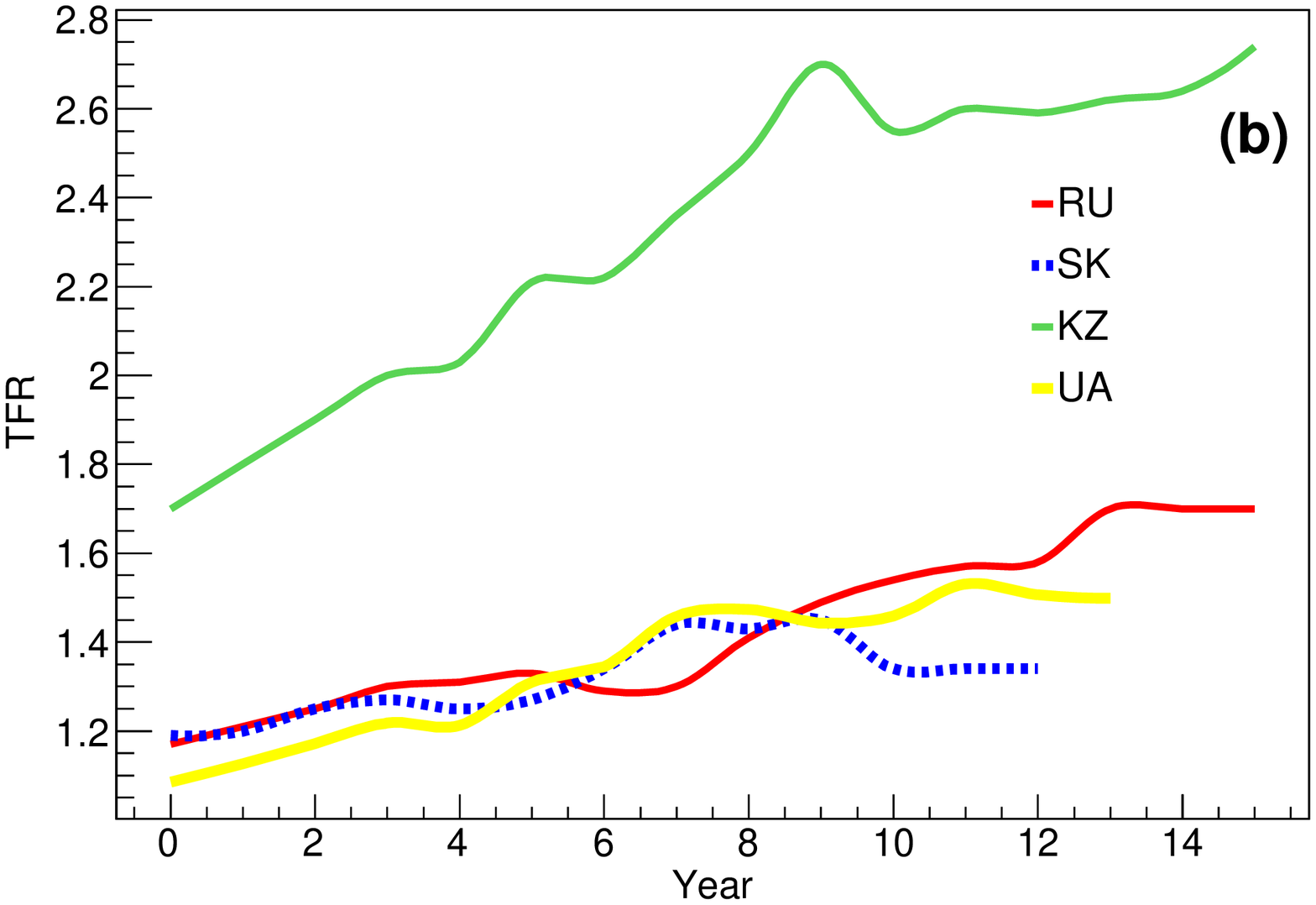}

\caption{Time series of the TFR from the year when the TFR reached its minimum, i.e., the reference year (Table \ref{tab:delta}), until 2014, in countries: (a) Vietnam (VN), Albania (AL), Azerbaijan (AZ) and Belarus (BY), and (b) Russian Federation (RU), Slovakia (SK), Kazakhstan (KZ) and Ukraine (UA). The corresponding differences of  $\Delta \mathrm{HDI}_{L}$ and $\Delta \mathrm{TFR}_{L}$ are shown in Table \ref{tab:delta}.}
\label{fig:refTFR}
\end{figure}

Myrskyl\"{a} \cite{Myrskyla} introduced a reference year as a year when the TFR is minimal. In Figure \ref{fig:refTFR}, time series of the TFR for years following the respective reference year are shown. Two trends can be seen  in Figure \ref{fig:refTFR}a, namely, a modest increase in Vietnam (VN) and  Albania (AL), and in Figure \ref{fig:refTFR}b) in the Russian Federation (RU), Slovakia (SK) and Ukraine (UA). Abrupt increase is shown for Belarus (BY) in Figure \ref{fig:refTFR}a  and for Kazakhstan (KZ) in Figure \ref{fig:refTFR}b).  We define  differences of the TFR and HDI as $\Delta\mathrm{TFR}_{L}=\mathrm{TFR}_{t}-\mathrm{TFR}_{r}$ 
and
$\Delta\mathrm{HDI}_{L}=\mathrm{HDI}_{t}-\mathrm{HDI}_{r}$, where $t$  is a year ($t=2014$), and $r$ is the reference year. The reference year, $\mathrm{HDI}_{r}$,  $\mathrm{HDI}_{t}$, $\Delta\mathrm{HDI}_{L}$, and $\Delta\mathrm{TFR}_{L}$ are summarised in Table \ref{tab:delta}.

Table \ref{tab:delta} shows countries (Country), country codes (Code), reference years (Year) \cite{Myrskyla},  $\mathrm{HDI}_{r}$ indices in  reference years $r$,  $\mathrm{HDI}_{t}$ indices in year $t=2014$, and corresponding changes of the HDI and TFR indices in years $t$ and  $r$, $\Delta \mathrm{HDI}_{L}$ and $\Delta \mathrm{TFR}_{L}$.

\begin{table}
\caption{\ The table of countries (Country), country codes (Code), reference years (Year), $\mathrm{HDI}_{r}$ in the reference year $r$, $\mathrm{HDI}_{t}$ in year $t=2014$, and their corresponding differences, $ \Delta \mathrm{HDI}_{L}$, and
$\Delta \mathrm{TFR}_{L}$. Countries where $\Delta \mathrm{TFR}_{L}>0$  are ordered in decreasing order of $\mathrm{HDI}_{r}$. }
\begin{center}
\begin{tabular}{  c  c  c  c  c  c  c  }
\hline
	Country & Code & Year & $\mathrm{HDI}_{r}$ & $\mathrm{HDI}_{t}$ & $ \Delta \mathrm{HDI}_{L}$  & $ \Delta \mathrm{TFR}_{L}$  \\ \hline
	Singapore & SGP & 2010 & 0.911 & 0.924 & 0.013 & 0.1 \tabularnewline
	Australia & AUS & 2001 & 0.902 & 0.937 & 0.035 & 0.12 \tabularnewline
	Switzerland & CHE & 2001 & 0.89 & 0.938 & 0.048 & 0.14 \tabularnewline
	Japan & JPN & 2005 & 0.873 & 0.902 & 0.029 & 0.16  \tabularnewline
	United Kingdom & GBR & 2001 & 0.87 & 0.908 & 0.038 & 0.2 \tabularnewline
	Canada & CAN & 2000 & 0.867 & 0.919 & 0.052 & 0.12 \tabularnewline
	Sweden & SWE & 1998 & 0.867 & 0.909 & 0.042 & 0.39 \tabularnewline
	New Zealand & NZL & 1998 & 0.863 & 0.913 & 0.05 & 0.03 \tabularnewline
	Korea Rep. & KOR & 2005 & 0.86 & 0.899 & 0.039 & 0.129  \tabularnewline
	Hong Kong SAR China & HKG & 2003 & 0.851 & 0.916 & 0.065& 0.333 \tabularnewline
	Slovenia & SVN & 2003 & 0.85 & 0.888 & 0.038 & 0.35 \tabularnewline
	Austria & AUT & 2001 & 0.847 & 0.892 & 0.045 & 0.11 \tabularnewline
	Germany & DEU & 1994 & 0.828 & 0.924 & 0.096 & 0.15 \tabularnewline
	Hungary & HUN & 2011 & 0.823 & 0.834 & 0.011 & 0.12 \tabularnewline
	Malta & MLT & 2007 & 0.813 & 0.853 & 0.04 & 0.03  \tabularnewline
	Czech Republic & CZE & 1999 & 0.811 & 0.875 & 0.064 & 0.33 \tabularnewline
	Spain & ESP & 1996 & 0.806 & 0.882 & 0.076 & 0.11 \tabularnewline
	France & FRA & 1993 & 0.803 & 0.894 & 0.091 & 0.26 \tabularnewline
	Israel & ISR & 1992 & 0.801 & 0.898 & 0.097 & 0.38 \tabularnewline
	Poland & POL & 2003 & 0.8 & 0.852 & 0.052 & 0.07  \tabularnewline
	Italy & ITA & 1995 & 0.799 & 0.881 & 0.082 & 0.2  \tabularnewline
	Greece & GRC & 1999 & 0.794 & 0.865 & 0.071 & 0.06 \tabularnewline
	Ireland & IRL & 1995 & 0.794 & 0.92 & 0.126 & 0.12 \tabularnewline
	Bahamas The & BHS & 2005 & 0.788 & 0.79 & 0.002 & 0.019 \tabularnewline
	Lithuania & LTU & 2002 & 0.78 & 0.846 & 0.066 & 0.36 \tabularnewline
	Slovak Republic & SVK & 2002 & 0.771 & 0.842 & 0.071 & 0.15 \tabularnewline
	Estonia & EST & 1998 & 0.758 & 0.863 & 0.105 & 0.24 \tabularnewline
	Lebanon & LBN & 2009 & 0.752 & 0.763 & 0.011 & 0.116 \tabularnewline
	Croatia & HRV & 1999 & 0.739 & 0.823 & 0.084 & 0.08 \tabularnewline
	Albania & ALB & 2008 & 0.721 & 0.762 & 0.041 & 0.159 \tabularnewline
	Trinidad and Tobago & TTO & 2001 & 0.721 & 0.779 & 0.058 & 0.034 \tabularnewline
	Romania & ROU & 2001 & 0.714 & 0.798 & 0.084 & 0.14 \tabularnewline
	Belarus & BLR & 2004 & 0.713 & 0.798 & 0.085 & 0.42 \tabularnewline
	Russian Federation & RUS & 1999 & 0.709 & 0.805 & 0.096 & 0.53 \tabularnewline
	Latvia & LVA & 1998 & 0.705 & 0.828 & 0.123 & 0.43 \tabularnewline
	Bulgaria & BGR & 1997 & 0.704 & 0.792 & 0.088 & 0.39 \tabularnewline
	Bosnia and Herzegovina & BIH & 2005 & 0.697 & 0.747 & 0.05 & 0.043 \tabularnewline
	Ukraine & UKR & 2001 & 0.683 & 0.748 & 0.065 & 0.413 \tabularnewline
	Georgia & GEO & 2001 & 0.677 & 0.768 & 0.091 & 0.224 \tabularnewline
	Kazakhstan & KAZ & 1999 & 0.676 & 0.793 & 0.117 & 1.04 \tabularnewline
	Tunisia & TUN & 2002 & 0.667 & 0.723 & 0.056 & 0.2 \tabularnewline
	Algeria & DZA & 2002 & 0.663 & 0.743 & 0.08 & 0.453 \tabularnewline
	Azerbaijan & AZE & 2001 & 0.651 & 0.758 & 0.107 & 0.2 \tabularnewline
	Moldova & MDA & 2005 & 0.648 & 0.701 & 0.053 & 0.037 \tabularnewline
	Egypt Arab Rep. & EGY & 2006 & 0.644 & 0.688 & 0.044 & 0.334 \tabularnewline
	Mongolia & MNG & 2002 & 0.609 & 0.733 & 0.124 & 0.574 \tabularnewline
	Vietnam & VNM & 2004 & 0.609 & 0.678 & 0.069 & 0.067 \tabularnewline
	Kyrgyz Republic & KGZ & 2000 & 0.593 & 0.662 & 0.069 & 0.8  \tabularnewline
	China & CHN & 2000 & 0.592 & 0.734 & 0.142 & 0.115 \tabularnewline
	Morocco & MAR & 2005 & 0.575 & 0.645 & 0.07 & 0.049 \\ \hline
\end{tabular}
\end{center}
\label{tab:delta}
\end{table}

\begin{figure}
\begin{center}
\includegraphics[width=7.5cm]{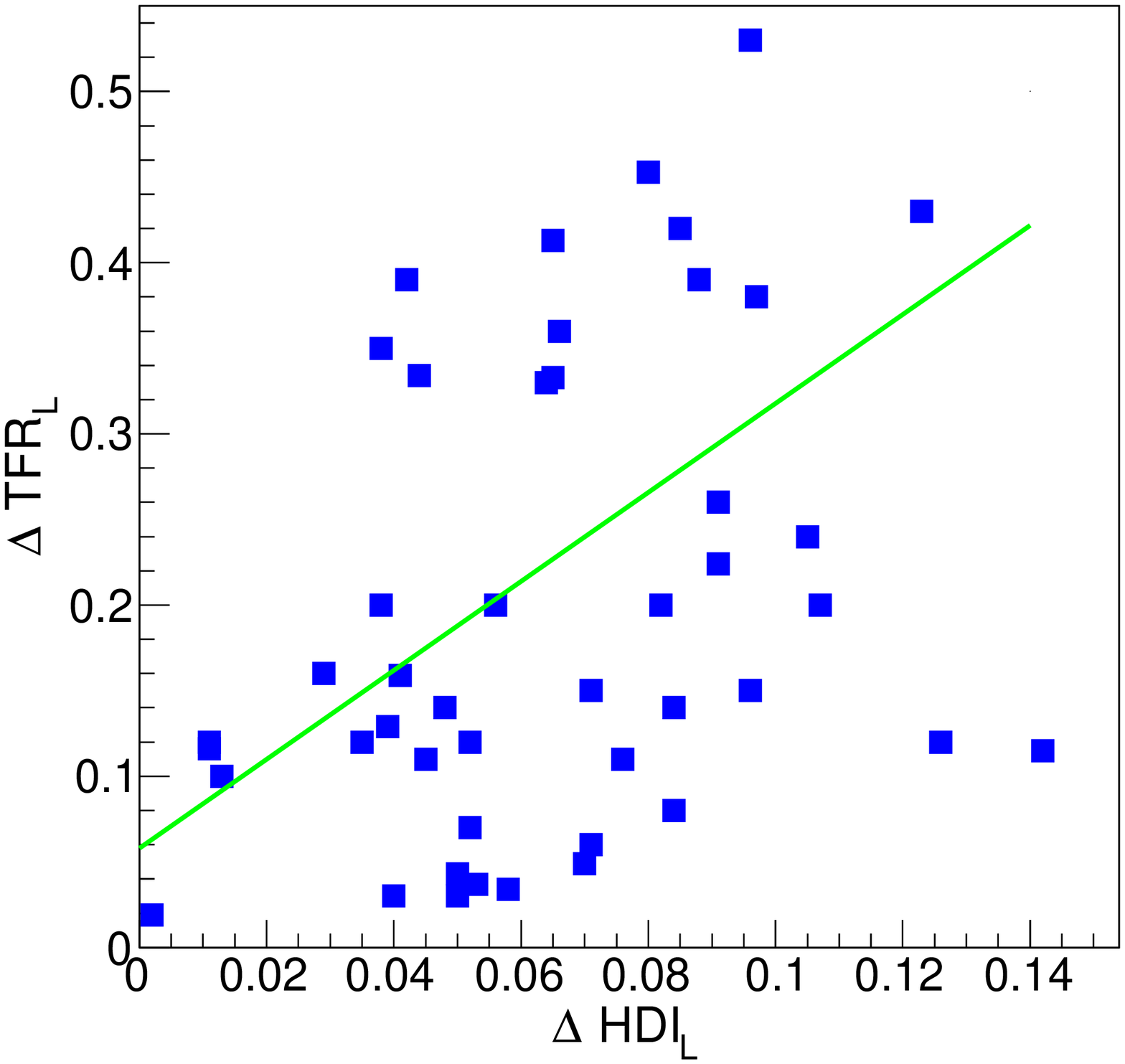}
\end{center}

\caption{ $\Delta\mathrm{HDI}_{L}$ versus $\Delta\mathrm{TFR}_{L}$ for the data in Table \ref{eq:DELTA}.  Spearman's rank coefficient for these data is $\rho=0.403 $, ($ P=0.004$) and the regression line shown is given by Eq. \ref{eq:DELTA}.}
\label{fig:correlat}
\end{figure}

The differences $\Delta\mathrm{HDI}_{L}$ and $\Delta\mathrm{TFR}_{L}$ are correlated (Figure \ref{fig:correlat}), where Speraman's rank coefficient is $\rho=0.403$,  ($ P=0.004$),  and regression line is  

\begin{equation}
\Delta\mathrm{TFR}_{L}=0.058+2.6\Delta\mathrm{HDI}_{L}.
\label{eq:DELTA}
\end{equation}

\section{Discussion}

The TFR began to decline in the United States in 2007, in Norway in 2009, and in Sweden in 2010  \cite{WBank}, and  the HDI simultaneously increased in these highly developed countries, as seen in  Table \ref{tab:TFRdecline}, which is an opposite trend to an expected increase of the TFR \cite{Myrskyla}. We associate this transient declining of the TFR in highly developed countries with the financial crisis of 2008.      

Correlations between the TFR and the HDI are important requirements to make clear conclusions about an impact of increase of the HDI on the TFR. In 2005, a new correlation interval,  $0.7 \leq \mathrm{HDI}\leq 1.0$ (Table \ref{tab:spearman}), with the rank correlation coefficient $\rho < 0$   emerged, and it persisted until 2014. However, Myrskyl\"{a} \textit{et al} \cite{Myrskyla} have not reported on this interval. Existence of this new correlation interval was the  first indication for us that the relationship between the HDI and the TFR could be more complex than claimed in Ref. \cite{Myrskyla}.    

We have analysed short time series of the TFR and the HDI (2010-2014) in the intervals  $\mathrm {HDI}_{i} < 0.85$, $0.7 \leq \mathrm{HDI}_{i}\leq 1.0$ and  $\mathrm{HDI}_{i} >0.85$, and the results (Table \ref{tab:spearman}) were compared  with the relation between the average $\langle\Delta \mathrm{TFR}_{k}\rangle$ and $\mathrm{HDI}_{C}$ shown in Figure \ref{fig:meas}, i.e., the HDI and the the TFR are not correlated for $\mathrm{HDI}>0.85$  due to an existence of both increasing and decreasing tendencies of the TFR in this interval when the HDI increases (Figure \ref{fig:meas}).

The rank coefficients (Table \ref{tab:spearman}) and the relation  $\langle\Delta \mathrm{TFR}_{k}\rangle$ versus $\mathrm{HDI}_{C}$ (Figure \ref{fig:meas}) support the empirical finding that the TFR decreases if the HDI increases, but a decrease  of the TFR for  $\mathrm{HDI>0.74}$ is lower than for  $\mathrm{HDI<0.74}$.  

Myrskyl\"{a} \textit{et al} \cite{Myrskyla}  found the TFR reversal at the critical value $\mathrm{HDI}_{C}=0.85 $ in 2005.  We found a transition from correlated to uncorrelated  relations between the TFR and the HDI in 2010-2014 in  the vicinity of this critical point,  $0.7 \leq \mathrm{HDI}\leq 0.85$ and  $\mathrm{HDI}\geq 0.85$ (Table \ref{tab:spearman}). These results of correlation analysis are supported by the measure $\langle\Delta \mathrm{TFR}_{k}\rangle$ vs $\mathrm{HDI}_{C}$ (Figure \ref{fig:meas}), where we can see the changes of the measure $\langle\Delta \mathrm{TFR}_{k}\rangle$ at $\mathrm{HDI}_{C}=0.82$ and $\mathrm{HDI}_{C}=0.87$.

We have found a new critical point  $\mathrm{HDI}_{CF}=0.74$ in Figure \ref{fig:meas} and in Table \ref{tab:spearman} that could be associated with a transition of the TFR  below the safe replacement level $\mathrm{TFR}=2.1$.  Societies below the safe replacement level are in danger of extinction \cite{Sneppen}. They have reached a new regime in which a rich spectrum of behaviours of the TFR could emerge, for example, the TFR will decline \cite{Kirk, Balter1894}, increase \cite{Myrskyla} or be a random value, i.e., there can be uncorrelated relation between the HDI and the TFR. Knowledge of the critical value of the HDI where the transition to below $\mathrm{TFR}=2.1$ takes place  could be important, however, more studies are needed to confirm this critical value. 

A temporal nature of the  TFR reversal in highly developed countries  is supported by time series of the TFR in these countries \cite{WBank}, our analyses of correlations of the TFR and the HDI (Table \ref{tab:spearman}), and statistical properties of differences of the TFR and the HDI (Figures \ref{fig:fig2} and \ref{fig:meas}). 

The HDI and the TFR  until 2009 show that a reversal in the HDI-TFR relationship is robust neither to the revision of HDI calculation method nor to the decomposition of the HDI into its subindices of education, health and living standard \cite{Harttgen_2014}; there is also also very little support for a simple interpretation that fertility rates will automatically start to increase beyond a certain level of development.     

The  HDI and the TFR  until 2014 (Section \ref{sec:reults}) confirm the existence of  turning points of the TFR reversal not only for highly developed countries, as Myskyl\"{a} \textit{el al.} found \cite{Myrskyla}, but also for medium developed counties with $0.575<\mathrm{HDI}<0.85$  (see Table \ref{tab:delta}). These results show that it is difficult to determine a single and universal transition value of the $\mathrm{HDI}_{C}$ where the fertility rate reversal takes place, and the relationship between the HDI and the TFR below the safe replacement level $\mathrm{TFR}=2.1$ is more complex than  was earlier reported \cite{Myrskyla}. 

The previous conclusions \cite{Myrskyla} that the TFR shows a reversal in highly developed countries   was confirmed only for data before the beginning of the financial crisis in 2008 \cite{Bongaarts_2012, luci_2010}, when economies worked well and unemployment rates were low \cite{Sobotka_2011}. The influences of economic crises on fertility trends in the developed world were systematically investigated in \cite{Sobotka_2011}, the authors  demonstrated the relationship between unemployment rate and total births (not the the TFR) in the case of Latvia (see Figure 2 in \cite{Sobotka_2011}).  The authors \cite{Myrskyla, luci_2010} considered  a tempo effect \cite{Sobotka_2011}  as a major effect that distorts the TFR, and they \cite{Sobotka_2011} explain a rise in the European fertility as disappearance of the tempo effect and socio-economic progress \cite{Myrskyla}. However, they \cite{Myrskyla,  Bongaarts_2012, luci_2010,Sobotka_2011} neglected the effect of continuous immigration \cite{Azose} that takes place during a long time from a periphery to the core \cite{Cernak2016}, and different economic regulations \cite{Cernak2016} that result in different TFR trends in, for example, USA, UK and Germany (Figure \ref{fig:TFrtimeser})\cite{Cernak2016}. 

In the previous work \cite{Cernak2016} we proposed a hypothesis that a long-term low $\mathrm{TFR} < 1.5$ is an unwanted state of a complex system, which is a state that could be a consequence of economic regulations to achieve sustained economic growth and low unemployment. The development of the TFR during an integration of Slovakia into the European Union and Eurozone (Figure \ref{fig:TFrtimeser}a, in 2000-2009) can serve as an example. Time series in the Czech Republic \cite{ Bongaarts_2012,Sobotka_2011} and Slovakia (Figure \ref{fig:TFrtimeser}a)  demonstrate  different trends of the TFR  before and after integration   of the Slovak republic  into Eurozone. During the period of 2004-2009, the Slovak national government regulated the society to fulfill criteria of the Exchange Rate Mechanism. The TFR in Slovakia was lower than in the Czech Republic: the TFR  in the Czech republic was the first time in history higher than in Slovakia (Figure \ref{fig:TFrtimeser}a). We note that Czechs and Slovaks separated in 1993.  Governments outside the Eurozone have much more freedom to influence their local economies and social status in the society, for example, in Sweden, Norway, UK, USA and Russia (Figure \ref{fig:refTFR}b). For more examples, see Figure 4 in \cite{luci_2010}.    
                
Sobotka \textit{et al} \cite{Sobotka_2011} consider recessions in the 1970s and economic shocks in Central and Eastern Europe after 1989 as different events. It is true that development in Central and Eastern Europe has some specific features \cite{Sobotka_2011}, however we believe that their impacts on the TFR decline are consequences of common reasons. We consider the TFR decline in the 1970s and 1990s as a response of a complex dynamical system \cite{Cernak2016} to new economic regulations, i.e. transition  from  the state of the safe $\mathrm{TFR}>2.1$ to the state with $\mathrm{TFR}<1.5$. The financial crisis in 2008 was specific because the transition took place in a different regime, i.e., many nations had TFR  below the safe TFR, and new economic and political tools  were used to achieve global economic growth. The derivatives $\frac{dTFR}{dt}$ (see Figure \ref{fig:TFrtimeser}) provide  details about the nature of the TFR decay  during the transition, as well as details about the background dynamics of the TFR.

The correlation of differences  $\Delta\mathrm{HDI}_{L}$ and $\Delta\mathrm{TFR}_{L}$ shown in Figure \ref{fig:correlat} is a new finding that provides an important information about a long-term dynamics of the TFR after the TFR reversal. An increase of the TFR is proportional to a change of the HDI, i.e., after the TFR reversal, it appears to be important to keep a certain economic and social progress to increase  the TFR in the long term. This conclusion is valid for many countries with different level of development $\mathrm{HDI}_{r}>0.575$ (see Table \ref{tab:delta}). However, it is not possible to increase the HDI infinitely, thus, in the global economy, migration from the peripheries to the core could be one, however only a temporal, solution to keep constant or growing population in the core. The main question is how sustainable such systems are. We assume that a systematic study of the correlation between  $\Delta\mathrm{HDI}_{L}$ and $\Delta\mathrm{TFR}_{L}$ and limits of the HDI increase after the turning point could have a big impact on social and economic changes in societies.

The low TFR and population decline are serious issues in the current economy, which is based on the growth of annual production. In this model the  global annual production, which is affected by the global consumption, needs to increase continuously. An increase of consumption can be achieved by increasing the number of consumers (population growth) or by increasing the individual consumption, however, both solutions have real limits.  Governments can control migration \cite{Azose}, support families \cite{Balter1894,Smeeding}, and support economies oriented on aging population \cite{Lee_2014, Elgin} to guarantee a continuous economic growth. The methods preferred by governments depend on economic optimizations that vary from country to country. The long-term impacts of such actions on the TFR and population size are difficult to predict \cite{Lee_2011, Azose} due to the complexity of these systems.


We also believe that it could be useful to consider  evolutionary principles \cite{Goodhart_1956,Burger2016, Bar-Yam} to better understand the nature of the TFR. 

The empirical relation between the TFR and the HDI \cite{Goodhart_1956, Myrskyla} and its universality in developing countries with $\mathrm{HDI}<0.74$ (see Figure \ref{fig:fig1} and Table \ref{tab:spearman}) could lead to a new strategy of international organizations to increase the HDI in these countries. Current strategies are focused only on two items of the HDI: better education of girls \cite{Gerland, Abel_2016} and higher quality of health care (prevention of HIV and other diseases). It is also necessary to enhance the third dimension of the HDI, namely, improve living standards (see definition of the  HDI \cite{UNDP}). 

We assume that societies in medium developed ($0.6<\mathrm{HDI}<0.85$) and highly developed ($\mathrm{HDI}>0.85$) countries  should be more active in searching for new economic and social models that will stimulate self-regulation of the TFR \cite{Goodhart_1956} around the safe replacement level \cite{Cernak2016}.

We assume that in the global world not only the level of the HDI is important, but also economic, political, and social changes that can significantly influence the individual choices of women to accept offspring \cite{Cernak2016}.  We think that the temporary break of the TFR reversal in 2010-2014 is a consequence of the financial crisis in 2007-2008 and the following economic and political actions. However, it is difficult to forecast the TFR trends.

\section*{Conflict of Interest Statement}

The author declares that the research was conducted in the absence of any commercial or financial relationships that could be construed as a potential conflict of interest.

\section*{Author Contributions}
The author confirms being the sole contributor of this work and approved it for publication.

\section*{Acknowledgments}
I thank V. Szappano\v{s}ov\'{a}, O. Smirnova and A. Read for reading the manuscript.

\bibliographystyle{unsrt} 
\bibliography{cernak}

\begin{thebibliography}{10}

\bibitem{Kirk}
D.~Kirk.
\newblock Demographic transition theory.
\newblock {\em Population Studies-a Journal of Demography}, 50(3):361--387,
  1996.

\bibitem{Dribe_2014}
Martin Dribe, Michel Oris, and Lucia Pozzi.
\newblock {Socioeconomic status and fertility before, during, and after the
  demographic transition: An introduction}.
\newblock {\em Demographic Research}, S14(7):161--182, 2014.

\bibitem{Goodhart_1956}
{Goodhart, C. B.}
\newblock {World Population Growth and its Regulation by Natural Means}.
\newblock {\em Nature}, 178(4533):561--565, sep 1956.

\bibitem{Lutz_2001}
W.~Lutz, W.~Sanderson, and S.~Scherbov.
\newblock The end of world population growth.
\newblock {\em Nature}, 412(6846):543--545, 2001.

\bibitem{Lee_2011}
R.~Lee.
\newblock The outlook for population growth.
\newblock {\em Science}, 333(6042):569--573, 2011.

\bibitem{Gerland}
P.~Gerland, A.~E. Raftery, H.~\v{S}ev\v{c}\'{i}kov\'{a}, N.~Li, D.~A. Gu,
  T.~Spoorenberg, L.~Alkema, B.~K. Fosdick, J.~Chunn, N.~Lalic, G.~Bay,
  T.~Buettner, G.~K. Heilig, and J.~Wilmoth.
\newblock World population stabilization unlikely this century.
\newblock {\em Science}, 346(6206):234--237, 2014.

\bibitem{Abel_2016}
G.~J. Abel, B.~Barakat, KC~Samir, and W.~Lutz.
\newblock Meeting the sustainable development goals leads to lower world
  population growth.
\newblock {\em Proceedings of the National Academy of Sciences},
  113(50):14294--14299, 2016.

\bibitem{Azose}
J.~J. Azose, H.~Sevcikova, and A.~E. Raftery.
\newblock Probabilistic population projections with migration uncertainty.
\newblock {\em Proceedings of the National Academy of Sciences of the United
  States of America}, 113(23):6460--6465, 2016.

\bibitem{Myrskyla}
M.~Myrskyl\"{a}, H.~P. Kohler, and F.~C. Billari.
\newblock Advances in development reverse fertility declines.
\newblock {\em Nature}, 460(7256):741--743, 2009.

\bibitem{Tuljapurkar}
S.~Tuljapurkar.
\newblock Demography babies make a comeback.
\newblock {\em Nature}, 460(7256):693--694, 2009.

\bibitem{Balbo}
N.~Balbo, F.~C. Billari, and M.~Mills.
\newblock Fertility in advanced societies: A review of research.
\newblock {\em European Journal of Population-Revue Europeenne De Demographie},
  29(1):1--38, 2013.

\bibitem{Burger2016}
O.~Burger and J.~P. DeLong.
\newblock What if fertility decline is not permanent? the need for an
  evolutionarily informed approach to understanding low fertility.
\newblock {\em Philosophical Transactions of the Royal Society B-Biological
  Sciences}, 371(1692), 2016.

\bibitem{Sneppen}
K.~Sneppen and G.~Zochi.
\newblock {\em Physics in Molecular Biology}.
\newblock Cambridge Univ. Press, New York, 2005.

\bibitem{Stulp}
G.~Stulp and L.~Barrett.
\newblock Wealth, fertility and adaptive behaviour in industrial populations.
\newblock {\em Philosophical Transactions of the Royal Society B-Biological
  Sciences}, 371(1692), 2016.

\bibitem{Bar-Yam}
Y.~Bar-Yam.
\newblock {\em Dynamics of Complex Systems}.
\newblock Addison- Wesley, Reading, MA, 1997.

\bibitem{Cho2009}
A.~Cho.
\newblock Ourselves and our interactions: The ultimate physics problem?
\newblock {\em Science}, 325(5939):406--408, 2009.

\bibitem{Cernak2016}
J.~Cernak.
\newblock The evolutionary approach to understand human low fertility
  phenomenon.
\newblock {\em Front. Phys. 5:11}, 2017.

\bibitem{Balter1894}
Michael Balter.
\newblock The baby deficit.
\newblock {\em Science}, 312(5782):1894--1897, 2006.

\bibitem{Cernak_1996}
J.~\v{C}ern\'{a}k.
\newblock Digital generators of chaos.
\newblock {\em Physics Letters A}, 214(3):151 -- 160, 1996.

\bibitem{Henson_2001}
Shandelle~M. Henson, R.~F. Costantino, J.~M. Cushing, Robert~A. Desharnais,
  Brian Dennis, and Aaron~A. King.
\newblock Lattice effects observed in chaotic dynamics of experimental
  populations.
\newblock {\em Science}, 294(5542):602--605, 2001.

\bibitem{WBank}
Data: The world bank.
\newblock {http://data.worldbank.org/indicator/}, 2016.
\newblock Used indicators: Fertility rate, total (births per woman) and
  Population, total. Accessed online on December 2016.

\bibitem{UNDP}
Human development data (1980-2015).
\newblock {http://hdr.undp.org/en/data}, 2016.
\newblock Used indicator: Human Development Index (HDI). Accessed online on
  December 2016.

\bibitem{Harttgen_2014}
Kenneth Harttgen and Sebastian Vollmer.
\newblock A reversal in the relationship of human development with fertility?
\newblock {\em Demography}, 51(1):173--184, Feb 2014.

\bibitem{Bongaarts_2012}
John Bongaarts and Tom\'{a}\v{s} Sobotka.
\newblock A demographic explanation for the recent rise in european fertility.
\newblock {\em Population and Development Review}, 38(1):83--120, 2012.

\bibitem{luci_2010}
Angela Luci and Olivier Thevenon.
\newblock {Does economic development drive the fertility rebound in OECD
  countries?}
\newblock working paper or preprint, September 2010.

\bibitem{Sobotka_2011}
Tom\'{a}\v{s} Sobotka, Vegard Skirbekk, and Dimiter Philipov.
\newblock Economic recession and fertility in the developed world.
\newblock {\em Population and Development Review}, 37(2):267--306, 2011.

\bibitem{Smeeding}
T.~M. Smeeding.
\newblock Adjusting to the fertility bust.
\newblock {\em Science}, 346(6206):163--164, 2014.

\bibitem{Lee_2014}
R.~Lee, A.~Mason, and N.~T.~A. Network.
\newblock Is low fertility really a problem? population aging, dependency, and
  consumption.
\newblock {\em Science}, 346(6206):229--234, 2014.

\bibitem{Elgin}
C.~Elgin and S.~Tumen.
\newblock Can sustained economic growth and declining population coexist?
\newblock {\em Economic Modelling}, 29(5):1899--1908, 2012.

\end{thebibliography}

\end{document}